Title :

Is it the first use of the word Astrobiology ?

Author :

Danielle Briot

Adress :

Observatoire de Paris

61 avenue de l'Observatoire

75014 Paris France

tel : 33(0)1 40 51 22 39 and 33(0)1 45 07 78 57

danielle.briot@obspm.fr

running title :

First use of the word Astrobiology ?

Abstract

The research of life in Universe is a ancient quest that has taken different forms over the centuries. It has given rise to a new science, which is normally referred as Astrobiology. It is interesting to research when this word was used for the first time and when this science developed to represent the search for life in Universe as is done today. There are records of the usage of the word "Astrobiology" as early as 1935, in an article published in a French popular science magazine. Moreover this article is quite remarkable because its portrayal of the concept of the subject is very similar to that considered today. The author of this paper was Ary J. Sternfeld (1905 - 1980), who was ortherwise known as a poorly respected great pioneer of astronautics. We provide a brief description of his life, which was heavily influenced by the tragic events of the 20th century history, from Poland and France to Russia. He was a prolific scientific writer who wrote a number of very successful scientific books and papers.

1. Introduction

The question of the life in the Universe, in relation with the question of the multiplicity of worlds, is very ancient and probably dates back to Greek philosophers. This question has taken many forms over the centuries. It is fascinating to research when and where it took its present form. It has given rise to a new science known by several names: astrobiology, or exobiology, or bioastronomy… However, when were the origins of this science established? When was this word with its current meaning first used? In this article, we attempt to provide answers to these questions.

First, we present the paper in which astrobiology was probably initially used and then provide a short biography of the author who was a quite unknown great scientist.

2. The paper

On its 1 July edition of 1935, the French popular science magazine, *La Nature*, which had been published since 1873, published a cover article, written in French of course, untitled *La vie dans l'Univers*, i.e. *Life in Universe*. As it is well known, very many studies and books had been written on this subject for several centuries. However, this paper written 77 years ago, has a form and presents the problem in a way very similar to present studies.

The first part of this article begins with an historical description starting from Greek philosophers and ending in the twentieth century. It covers the range of beliefs, representations, and hypotheses about life in the Universe, and critically summarises the studies of Mars canals in which the author appears to disbelieve. The second part of the article reviews the current research situation in the problem of life in Universe. The section of the paper entitled *Origin of life* begins "*The development of both the natural and astronomical* *sciences has led to birth of a new science whose main objective is to assess the habitability of the other worlds, this science is called astrobiology.*". This is one of the oldest and may be the oldest, use and definition of the word *astrobiology* with its present meaning. The following sentence is : "*A comprehensive definition of life does not exist as far as we know, hence we leave this to the reader's intuition.*". This sentence is followed by a review of the various theories of the origin of life, and a study about the transplantation of life, corresponding to the panspermia. In the following paragraph, the biological point of view is reviewed, defining some extreme external conditions. The probability that some planets orbit other stars is studied according to the solar system formation model, model of Kant-Laplace or near-collision hypothesis of Jeans. In the model of Kant-Laplace: "*Owing to the infinity of planetary systems, the probability of finding some celestial objects where comptable with life, would be very high.*", and in the model of Jeans, the probability of finding other planetary systems is really low. The article also reviews the properties of the atmospheres of various planets and states that Titan, the largest satellite of Saturne, probably has an atmosphere. This prediction was remarkable in its correction. Let us recall that in 2005, the Huygens probe, detached from the Cassini spacecraft (a NASA-ESA space mission), landed on Titan to analyze its atmosphere and research hypothetical traces of life.

The conclusion of this article is : "*Our main conclusion is that all possibilities remain open and that nothing has been proved. Despite the considerable effort of famous astronomers, and advances in astronomical instrumentation, the question of whether life on other planets exist remains unanswered. Can we ever hope to reach any conclusion ?* "

This article written by Sternfeld is particularly remarquable because it took account the very last scientific discoveries, presented many original ideas, and was far removed from any imaginative extrapolation. We have to keep in mind that this article was written 77 years ago. The time interval between the publication of this paper and the present time is quite the same as between the publication of the *Sidereus Nuncius* by Galileo Galilei (1610) and *Conversations on the Plurality of Worlds* by Fontenelle (1686), whereas the presentation of the question of research of life in the Universe in this article, is quite similar to current ones at the present time.

As said above, this article contains one of the oldest and maybe the oldest use and definition of the word *astrobiology* with its modern-days meaning. A few years later, Lafleur (1941) defined the word *astrobiology* as "*the consideration of life elsewhere than on Earth*" in a article entitled *astrobiology*, and this is often considered as the oldest use of the word *astrobiology*. In 1945, Tikhov, a Russian astronomer, coined the word *astrobotany* that he used for the search of a vegetation on Mars. He used at the same time the word *astrobiology*, considering that he is one of the first persons to use this word, and a little later he used the word *cosmobiology* as a generalization for research about life on other planets, and particularly planets orbiting around other stars than our Sun. The first American symposium in astrobiology was held in 1957 (Wilson, and following papers, 1958), but the sense of the word *astrobiology* was not as restricted as its present meaning, and the papers of this symposium concerned not only life in other celestial bodies, but also all the problems common to astronomy and biology, among them, for example, physiological problems of astronomical observations. In 1965, Mamikunian attributed to Joshua Lederberg (1960) the creation of the word *exobiology*. A more detailed study of the use of the word *astrobiology* is given in another paper (Briot, 2012)

3.The author of the article: Ary Sternfeld

The article *La vie dans l'Univers* i.e. *Life in the Universe*, is so modern, so interesting and was so ahead of its time and relevant to research performed today, so that we wish to know what was his author. This paper is signed by: *Ary J. Sternfeld - prix international d'astronautique* i.e. International Astronautics Award. Actually, this author was no ordinary scientist and several studies have described his life. He is widely considered to be a pioneer of astronautics, even though he remained unrecognised as such form many years. His personnality, his life and his work are really outstanding and deserves to be known. Some information can be obtained about his life in one of his papers, (Sternfeld, 1965), or in a paper by Lucius (1981), and particularly in the book *From astronautics to Cosmonautics* written by Gruntman (2007). However, no-one has ever reviewed his paper on astrobiology which is why it is the subject of the current review.

As I describe below, his life was spent in several different countries, where his name took various forms: Ary Jacob Sternfeld, Arje Jakob Stzernfeld, Ary Abramovič Šternfel'd, Ario Abramovich Shternfel'd, Arii Abramovich Shternfel'd, Ary Chternfeld, Arieh Sternfeld, Штернфельд..... The word Sternfeld in Yiddish, as well as in German, means Star Field. We provide a description of his life and work, particularly the period during which he wrote his article about life in the Universe.

Ary Sternfeld was born in 1905, in Sieradz, a little town in Poland, in a Jewish family. From a young boy, he was strongly interested in astronautics and space flights. Because of a *numerus clausus* for Jewish students, he could not attend the Polytechnical School in Warsaw. He was a student at the Jagellon University in Krakov. He came to France, which represented the Land of Freedom, in 1924. His life as a student in France is described in one of his papers (Sternfeld 1965). To begin with, he spent several months in Paris to earn money to finance his much-hoped-for years of study at a university. He initially worked in a food market and then in various factories in both Paris and its suburbs. In October 1924, he went to Nancy, in the East of France, where he was accepted as a engineering student at the Mechanical Institute of the Nancy University. Because he did not speak French, he had a very difficult time in the beginnings, but soon made rapid progress. At the end of both his third and final year at Nancy, he was second place in his class. His living conditions as a student were incredibly harsh. He could eat only the cheapest of food and worked throughout the university holidays to earn money to live. After graduating from Nancy in 1927, he returned to Paris, where he again worked in various factories but as an engineer, and machine designer. He applied to do a doctorate in astronautics at the Sorbonne, Science University in Paris, but his proposed thesis subject was rejected because he was far too ahead of his time ! We emphasize that this was indeed a long time before the first spacecraft, the first artificial satellite, Sputnik, was launched in October 1957. Sternfeld visited Russia for one month in 1932, then returned to Poland for one year. There, he wrote a book in French: *Initiation à la Cosmonautique* i.e. *Introduction to Cosmonautics*. In this book, he coined the word *Cosmonautics*, to be used instead of the word *Astronautics*, which itself was coined by the French Robert Esnault-Pelterie (see Gruntman, 2005). His 490-page manuscript was ready in November 1933. Among many other things, this book contains some calculations of the most relevant type of space velocities. A long time afterwards, in 1962, Sternfeld published a paper in French in the Bulletin de la Société Lorraine des Sciences, edited in Nancy, the city where he has been a student and graduated (Sternfeld, 1962). In this paper, he reviewed the ideas and calculations of trajectories for spacecrafts that were prescient in this book. He concluded that the differences between theoretical trajectories calculated more than 25 or 30 years in advance, and effective ones for the first four artificial satellites are equal to 1% on average. One of the orbits he calculated is now named the Sternfeld orbit. He returned to France with his manuscript in December 1933. In 1934, he sent two notes at the French Science Academy (Sternfeld 1934a, 1934b). He won the International Astronautic REP (Robert Esnault-Pelterie) Hirsch Award of Encouragement. From December 1933, to June 1935, that is to say one year and half, he published a number of articles in popular and professional publications in France, at least 6 in 1934 and 7 in 1935, among them his article about the life in the Universe. A list of at least some of these publications is given in the book of Gruntman (2005). Some months before his article about life in the Universe appeared, he published an article in another issue of the same popular science magazine, entitled *Signalisation Interplanétaire*, i.e. *Interplanetary Signals*, about the possible ways for communicating between planets, using luminous signals reflected by mirrors or using radio waves (Sternfeld 1935a). As reported by Gruntman (2005), he said also that he was publishing a novel within a weekly magazine. We have now to research this magazine and this novel !

Nevertheless, he perceived that there was insufficient interest in France for astronautics. He and his wife were communists and the Soviet Union was therefore appealing to them particularly when a position there was offered to Sternfeld. They went to Russia in June 1935, and Ary Sternfeld joined the Jet Propulsion Scientific Research Institute. While he was working here, he also prepared the translation in Russian of his manuscript *Initiation à la Cosmonautique*. His scientific activities in Soviet Union remained undisclosed in France for some time. In accordance with a paper published by the director of the Nancy Electricity and Mechanics Institute (Lucius, 1981) after his death, it is reasonable to assume that nothing was ever revealed because it was always a state secret, and also because there was little communication from Soviet Union to Western countries. The reality, however, was very different. Since Sternfeld had graduated from a French university, had won a French research prize, both in a capitalist country, and perharps also was a Jew, he was dismissed in 1937 from his institute, and in 1938 from an other institute where he was working temporarily. After that, he was never able to work in any Russian institute specialising in space technology. It is possible and even probable that this apparently unfortunate event saved his life because a few weeks later, Stalin's infamous purge of scientists occured, during which scientists of this institute were arrested, deported in camps and very often shot. Afterwards, Sternfeld worked at home and continued with his research, writing many books and papers. He wrote in total around 30 books and over 400 popular and scientific articles. His books were very successful, specially after the launch of Sputnik, the first artificial satellite, in October 1957. His books were published in 39 countries, and translated into 36 languages. He ough to have become a rich man, but since the Soviet Union had never signed any international agreement about author copyright, Sternfeld was unable to gain royalties for any book published abroad. He was able to travel in Poland several times because Poland was a Socialist country behind the Iron Curtain, but was never able to obtain a visa allowing to return to France, even when he was awarded an honorary doctorate by Nancy University in 1961, or when he won the Galabert Astronautic International Prize in 1962. He died in 1980, while Nancy University continue to hope that he could be granted a visa to return to France.

He had a very hard life, but, by a kind of miracle, he survived the Stalin's repression in Russia. Nobody could know what might have happened if Sternfeld had remained in Poland as a Jew, or in France as a Polish Jew, during the Second World War and the Nazi occupation, considering the Nazi program to exterminate Jewish people.

A crater on the far side of the Moon bares his name, the planetarium and observatory in Lotz was named after him, as well as a *people museum* in a Russian boarding school for children with impaired hearing. A special tribute was paid to him when the NASA New Horizons spacecraft, was launched in 2006, towards Pluto. A small plaque onboard was engraved with the names of some pioneers of spaceflight mechanics and among them is Ary Sternfeld (Gruntman, 2007).

Ary Sternfeld is considered as a great pioneer in astronautics, and should equally be regarded as an incredibly original pioneer of Astrobiology.

4. Conclusion

It is fascinating to research the history of astrobiology because many studies and papers which containing many original ideas continue to remain unexplored. The article of Ary J. Sterfeld published as soon as 1935 is a good illustration of this. The main ideas behind the article, which are similar to those of present studies, are remarkably innovative for the time. However, we can presume that future studies about the Life in Universe will have a different form and will show real progress from study to study, when the answer to the question: "*Is there some life outside the Earth ?*" will be obtained. Then the history of astrobiology will really begin.

**Acknowledgements.**

Many thanks are due to Jean Schneider and Florence Raulin-Cerceau for very helful and encouraging discussions, and to Claire Halliday for her very careful reading of the manuscript.

**Author Disclosure Statement**

No competing financial interests exists.

**References**

Briot, D. (2012) Evolution of the problem "Research of life in the Universe" from some examples. Workshop Drake Formula. *To be published*

Fontenelle, B. le Bouyier de. (1686) *Entretien sur la Pluralité des Mondes*. Paris: Vve. C. Blageart.

Galilei, G. (1610) *Sidereus Nuuncius*. Venice: Thomas Baglioni.

Gruntman, M. (2007) From Astronautics to Cosmonautics. Publisher: *BookSurge*, LLC, North Charleston, South Carolina.

Lafleur, L. J. (1941) Astrobiology. *Astronomical Society of the Pacific Leaflets,* Vol. 3, No. 143, pp. 333-340.

Lederberg, J. (1960) Exobiology: approaches to life beyond the Earth. *Science*, 132:393-400.

Lucius M., (1981) Ary Sternfeld 1905 – 1980. *Bulletin Académie et Société Lorraines des Sciences*, Tome XX - n°2 – pp. 35-50.

Mamikunian, G. (1965), in Current aspects of Exobiology, ed. G. Mamikunian & M. H. Briggs, *Pergamon Press*, p.viii.

Sternfeld A.J. (1934a) Méthode de détermination de la trajectoire d'un corps en mouvement dans l'espace interplanétaire par un observateur lié au système mobile. *Comptes Rendus de l'Académie des Sciences*, Paris, 198:333-334.

Sternfeld A.J. (1934b) Sur les trajectoires permettant d'approcher d'un corps attractif central à partir d'une orbite keplérienne donnée. *Comptes Rendus de l'Académie des Sciences, Paris*, 198:711-713.

Sternfeld, A. J. (1935a) Signalisation Interplanétaire. *La Nature*, Masson et Cie Eds, Paris, N◦ 2944: 1-7.

Sternfeld, A. J. (1935b) La vie dans l'Univers. *La Nature*, Masson et Cie Eds, Paris, N◦ 2956, pp1-12.

Sternfeld, A. (1962) Idées prioritaires en astronautique. *Bulletin de la Société Lorraine des Sciences*, Tome II n°1, 28-33.

Sternfeld, A. (1965) En France, hospitalière et généreuse. *Grandes écoles à Nancy*, N°12: 96-101.

Wilson, A.G. (1958) Problems common to the fields of astronomy and biology: a symposium – Introduction. *Publications of the Astronomical Society of the Pacific*, 70:41-43.